\documentclass[a4paper]{article}

\begin{document}


\def\simlt{\stackrel{<}{{}_\sim}}
\def\simgt{\stackrel{>}{{}_\sim}}

\newcommand{\lsim}{\mbox{\raisebox{-.9ex}{~$\stackrel{\mbox{$<$}}{\sim}$~}}}
\newcommand{\gsim}{\mbox{\raisebox{-.9ex}{~$\stackrel{\mbox{$>$}}{\sim}$~}}}

\newcommand\vev[1]{{\langle {#1} \rangle}}

\renewcommand\({\left(}
\renewcommand\){\right)}
\renewcommand\[{\left[}
\renewcommand\]{\right]}

\newcommand\del{{\mbox {\boldmath $\nabla$}}}

\newcommand\eq[1]{Eq.~(\ref{#1})}
\newcommand\eqs[2]{Eqs.~(\ref{#1}) and (\ref{#2})}
\newcommand\eqss[3]{Eqs.~(\ref{#1}), (\ref{#2}), and (\ref{#3})}
\newcommand\eqsss[4]{Eqs.~(\ref{#1}), (\ref{#2}), (\ref{#3})
and (\ref{#4})}
\newcommand\eqssss[5]{Eqs.~(\ref{#1}), (\ref{#2}), (\ref{#3}),
(\ref{#4}) and (\ref{#5})}
\newcommand\eqst[2]{Eqs.~(\ref{#1})--(\ref{#2})}

\newcommand\eqref[1]{(\ref{#1})}
\newcommand\eqsref[2]{(\ref{#1}) and (\ref{#2})}
\newcommand\eqssref[3]{(\ref{#1}), (\ref{#2}), and (\ref{#3})}
\newcommand\eqsssref[4]{(\ref{#1}), (\ref{#2}), (\ref{#3})
and (\ref{#4})}
\newcommand\eqssssref[5]{(\ref{#1}), (\ref{#2}), (\ref{#3}),
(\ref{#4}) and (\ref{#5})}
\newcommand\eqstref[2]{(\ref{#1})--(\ref{#2})}

\newcommand\pa{\partial}
\newcommand\pdif[2]{\frac{\pa #1}{\pa #2}}

\newcommand\ee{\end{equation}}
\newcommand\be{\begin{equation}}
\newcommand\eea{\end{eqnarray}}
\newcommand\bea{\begin{eqnarray}}

\newcommand\mpl{M_{\rm P}}

\newcommand\dbibitem[1]{\bibitem{#1}\hspace{1cm}#1\hspace{1cm}}
\newcommand{\dlabel}[1]{\label{#1} \ \ \ \ \ \ \ \ #1\ \ \ \ \ \ \ \ }
\def\dcite#1{[#1]}

\def\calf{{\cal F}}
\def\calh{{\cal H}}
\def\call{{\cal L}}
\def\calm{{\cal M}}
\def\caln{{\cal N}}
\def\calp{{\mathcal P}}
\def\calr{{\cal R}}
\def\calpr{{\calp_\calr}}

\newcommand\bfa{{\mathbf a}}
\newcommand\bfb{{\mathbf b}}
\newcommand\bfc{{\mathbf c}}
\newcommand\bfd{{\mathbf d}}
\newcommand\bfe{{\mathbf e}}
\newcommand\bff{{\mathbf f}}
\newcommand\bfg{{\mathbf g}}
\newcommand\bfh{{\mathbf h}}
\newcommand\bfi{{\mathbf i}}
\newcommand\bfj{{\mathbf j}}
\newcommand\bfk{{\mathbf k}}
\newcommand\bfl{{\mathbf l}}
\newcommand\bfm{{\mathbf m}}
\newcommand\bfn{{\mathbf n}}
\newcommand\bfo{{\mathbf o}}
\newcommand\bfp{{\mathbf p}}
\newcommand\bfq{{\mathbf q}}
\newcommand\bfr{{\mathbf r}}
\newcommand\bfs{{\mathbf s}}
\newcommand\bft{{\mathbf t}}
\newcommand\bfu{{\mathbf u}}
\newcommand\bfv{{\mathbf v}}
\newcommand\bfw{{\mathbf w}}
\newcommand\bfx{{\mathbf x}}
\newcommand\bfy{{\mathbf y}}
\newcommand\bfz{{\mathbf z}}

\newcommand\yr{\,\mbox{yr}}
\newcommand\sunit{\,\mbox{s}}
\newcommand\munit{\,\mbox{m}}
\newcommand\wunit{\,\mbox{W}}
\newcommand\Kunit{\,\mbox{K}}
\newcommand\muK{\,\mu\mbox{K}}

\newcommand\metres{\,\mbox{meters}}
\newcommand\mm{\,\mbox{mm}}
\newcommand\cm{\,\mbox{cm}}
\newcommand\km{\,\mbox{km}}
\newcommand\kg{\,\mbox{kg}}
\newcommand\TeV{\,\mbox{TeV}}
\newcommand\GeV{\,\mbox{GeV}}
\newcommand\MeV{\,\mbox{MeV}}
\newcommand\keV{\,\mbox{keV}}
\newcommand\eV{\,\mbox{eV}}
\newcommand\Mpc{\,\mbox{Mpc}}

\newcommand\msun{M_\odot}

\newcommand\sub[1]{_{\rm #1}}
\newcommand\su[1]{^{\rm #1}}

\newcommand{\one}{_1}
\newcommand{\two}{_2}

\newcommand\mone{^{-1}}
\newcommand\mtwo{^{-2}}
\newcommand\mthree{^{-3}}
\newcommand\mfour{^{-4}}
\newcommand\mhalf{^{-1/2}}
\newcommand\half{^{1/2}}
\newcommand\threehalf{^{3/2}}
\newcommand\mthreehalf{^{-3/2}}

\renewcommand{\P}{{\cal P}}
\newcommand{\f}{f_{\rm NL}}
\newcommand{\mk}{{\mathbf k}}
\newcommand{\mq}{{\mathbf q}}

\newcommand{\zetag}{{\zeta\sub g}}
\newcommand\sigmas{{\sigma^2}}

\newcommand{\fnl}{{f\sub{NL}}}
\newcommand{\fnltilde}{{\tilde f\sub{NL}}}
\newcommand{\fnli}{ {f_{ {\rm NL}i } } }

\newcommand\kmax{{k\sub{max}}}
\newcommand\bfkp{{{\bfk}'}}
\newcommand\bfkpp{{{\bfk}''}}
\newcommand\bfpp{{{\bfp}'}}

\newcommand\tp{{(2\pi)}}
\newcommand\tpq{{(2\pi)^3}}
\newcommand\tps{{(2\pi)^6}}

\renewcommand\ni{{N_{,i}}}
\newcommand\nj{{N_{,j}}}
\newcommand\nij{{N_{,ij}}}
\newcommand\nii{{N_{,ii}}}

\title{The bound $r\leq16\epsilon$ on the primordial tensor perturbation}

\author{David H. Lyth\\
Department of Physics, Lancaster University, Lancaster LA1 4YB, UK}
 
\date{}

\maketitle

\begin{abstract}
I  recall  the well-known sufficient conditions for the bound 
 $r<16\epsilon$ on the spectrum of the 
primordial tensor perturbation. Two  recent papers claim a violation of this
bound, without stating explicitely any violation of the sufficient conditions.
\end{abstract}

The following conditions are sufficient for the bound $r<16\epsilon$
on the primordial tensor perturbation.
\begin{itemize}
\item
Inflation is of the slow-roll variety.
\item 
The  curvature perturbation is generated from the vacuum fluctuation  of
one or more light scalar fields.
\item
Einstein gravity holds during and after inflation.
\end{itemize}
Light fields are here defined as those whose vacuum fluctuations generate
classical  perturbations a few Hubble times after horizon exit,
which are uncorrelated and have a spectrum well-approximated by  $(H/2\pi)^2$.
According to the second condition, these perturbations provide the initial
condition for the evolution of the curvature perturbation.

Taking into account also the first condition,   the spectrum of the 
curvature perturbation is of the form
\be
\calp_\zeta(k,t)  = \( \frac{H^2}{2\pi\dot\phi} \)^2 + \cdots
\label{1}
.
\ee
The first term is the contribution of the perturbation in the inflaton field 
 $\phi$,  defined in this context as
the one pointing along the inflaton trajectory soon after  relevant scales 
leave the horizon. It is to be evaluated at horizon exit, making it 
time-independent. The 
 dots indicate the  possible  contributions of light fields 
 orthogonal to the inflaton field.
These are positive and are
initially negligible compared with the  one from the inflaton field, 
but they could grow to become
the dominant contribution. We are interested in the spectrum when it has 
levelled out to its final value constant value,  which according to 
observation is achieved by the time that the  Universe is a few seconds
old.

The above result was given by Starobinsky \cite{starob}
 and by Sasaki and Stewart \cite{ss},
 who had in mind that  $\zeta$ would achieve its final value 
by the end of inflation. It was
given again by Lyth and Riotto \cite{treview}
 who mentioned the possibility that
$\zeta$ may achieve its final value only after inflation is over. 
Among possible realizations of the latter case are the curvaton scenario 
\cite{curvaton,earlier}.

The third condition gives the spectrum of the primordial tensor perturbation.
Adopting a common convention for its normalization and setting $8\pi G=1$
the spectrum  is 
 \be
\calp\sub t(k) = 8(H/2\pi)^2
\label{2}
,
\ee
where the right hand side is to be  evaluated at horizon exit.
Combining \eqs{1}{2} and introducing the slow-roll parameter
 $\epsilon= \frac12 \dot\phi^2/H^2$ gives the advertised bound
\cite{ss}.

This note, which will not be submitted for publication in a journal,
 was prompted by the appearance of two papers
\cite{bkr,sloth}. They claim a violation of  the bound without stating 
explicitly any violation of the three sufficient conditions.

\end{document}